\documentclass[12pt]{article}

\usepackage{amsmath,amssymb,theorem}
\usepackage{fullpage,epsfig}

\newtheorem{thm}{Theorem}[section]
\newtheorem{lemma}[thm]{Lemma}
\newtheorem{prop}[thm]{Proposition}

{\theorembodyfont{\rmfamily}
\newtheorem{defn}[thm]{Definition}

\newtheorem{rmk}[thm]{Remark}}

\newcommand{\qed}{\hfill \mbox{\raggedright \rule{.07in}{.1in}}}
 
\newenvironment{proof}{\vspace{1ex}\noindent{\bf
Proof}\hspace{0.5em}}{\hfill\qed\vspace{1ex}}

\newcommand{\R}{{\mathbb R}}

\newcommand{\clos}{\operatorname{clos}}

\renewcommand{\O}{{\bf O}}
\newcommand{\SE}{{\bf SE}}
\newcommand{\E}{{\bf E}}

\begin{document}

\title{Dynamics on unbounded domains; 
\\ co-solutions and inheritance of stability}

\author{Peter Ashwin\\
Mathematics Research Institute\\
University of Exeter\\
Exeter EX4 4QE, UK
\and
Ian Melbourne\\
Department of Mathematics and Statistics\\
University of Surrey \\
Guildford GU2 7XH, UK
}

\date{\today}
\maketitle

\begin{abstract}
We consider the dynamics of semiflows of patterns on unbounded domains 
that are equivariant under a noncompact group action. We exploit the unbounded 
nature of the domain in a setting where there is a strong `global' norm and
a weak `local' norm. Relative equilibria whose group orbits are closed manifolds for a 
compact group action need not be closed in a noncompact setting;
the closure of a group orbit of a solution can contain `co-solutions'. 

The main result of the paper is to show that co-solutions inherit stability 
in the sense that co-solutions of a Lyapunov stable pattern are also stable (but 
in a weaker sense). This means that the existence of a single group orbit 
of stable relative equilibria may force the existence of quite 
distinct group orbits of relative equilibria, and these are also stable. This is in 
contrast to the case for finite dimensional dynamical systems where group orbits
of relative equilibria are typically isolated.
\end{abstract}

\section{Introduction}

There has been much effort devoted to trying to understand the dynamics of
spatially extended systems; that is, dynamical systems that have not just unbounded
time but also unbounded space dependence.
Most of this work has progressed by restricting to parabolic 
partial differential equations such  as the Ginzburg-Landau 
and Swift-Hohenberg equations (see for example \cite{Mielke,Fei&al96})
used to model `generic' instabilities with nontrivial
spatial dependence. This has been successful in characterising solutions
of specific types in specific systems of equations; for
example wave-like solutions, fronts between them \cite{Col&Eck90}, spirals and defects.
Related to this approach there have been attempts to find
a `qualitative theory' of partial differential equations (see for example
\cite{CouRieTre00}) where careful reductions to ODE models can explain many
universal features of patterns in unbounded systems, for example the stability
of fronts and defects \cite{SanSch}.

There remain fundamental problems in trying to characterise what sort of
attractors `generically' appear in initial value problems for partial 
differential equations with unbounded domains. Related to this is the 
usual problem of deciding
which space of functions and which topology or norm is appropriate. As 
is well recognised, a change of choice of norm can lead to qualitatively very
different behaviour \cite{Mie&Sch95,Mielke}. For example, consider a solution $u(x,t)$
to the heat equation $u_t=u_{xx}$ 
with initial condition $u_0(x)\geq 0$ satisfying $\int |u_0(x)|\,dx<\infty$.
Then $u(x,t)$ decays to zero in sufficiently `weak' norms; for example
a weighted $L^1$ norm
$\|u\|= {\textstyle\int} \rho(x) |u(x,t)|\,dx$ with weight $\rho>0$ 
such that $\rho(x)\rightarrow 0$ as $|x|\rightarrow \infty$.
By contrast in a more `global' norm such as the $L^1$ norm, the solution 
remains bounded away from zero.

Rather than being discouraged by what might be seen as the arbitrary 
nature of the choice of topology, in this paper we wish to use it to our 
advantage. In an attempt to move away from specific systems of 
equations we consider a general setting where the dynamics is given 
by a semiflow on a space of patterns. A key assumption 
is that there are two important topologies;  a weak one which characterises 
local changes and a strong one which characterises global changes. We 
assume that the semiflow is continuous for both of the topologies and note that 
this is typical in many systems. 

If there is a stable propagating front between states $A$ and $B$ it seems reasonable to
ask whether $A$ and $B$ inherit the stability of the front. We describe a setting in which 
one can make such deductions and extend them to understand the stability in
general for `far field' stable patterns.

In many models for dynamics on unbounded domains there are 
translational or Euclidean group symmetries in the model. We make 
an assumption of this form to allow us to discuss a range of different 
solutions. We assume that the semiflow commutes with the action
of a noncompact group $\Gamma$, meaning that we can characterise the 
unboundedness by group symmetries.

Section~\ref{secsetting} gives the details of the setting and some motivating examples.
Section~\ref{seccosolutions} shows that under quite general assumptions on the semiflow, 
solutions often force the existence of a variety of new solutions
that we call {\em co-solutions}. These co-solutions are patterns that are
in the closure of the group orbit in the weak topology. In Proposition~\ref{prop-cosols}
we give conditions such that if the original solution is a relative equilibrium then 
the co-solution is also a relative equilibrium. 
The main result in Section~\ref{secinheritproof} shows that the stability of a solution
implies stability of its co-solutions. Section~\ref{secdiscuss} discusses 
the results and suggests some further questions that may be usefully addressed using 
this approach.

\section{Semiflows with noncompact domain symmetries}\label{secsetting}

We consider the behaviour of patterns that evolve on an infinite 
domain for a semiflow equivariant under a group $\Gamma$ acting 
on this domain. By a space of {\em patterns} we mean a 
vector space $\mathcal{B}$ of functions $u:D\to\R^s$ where for example 
$D=\R^n$. We suppose that $\Gamma$ is a (noncompact) Lie group 
acting on $\mathcal{B}$ and $L\Gamma$ is the Lie algebra of $\Gamma$.
Our main applications have $D=\R^n$ and $\Gamma=\E(n)$ or $\SE(n)$.

\subsection{The strong and weak topologies}

We consider semiflows on $\mathcal{B}$ subject to two topologies
that express closeness in a weak (local) or in a strong (global) senses.
More precisely, we assume:

\begin{itemize}
\item[(H1)]
\begin{itemize}
\item[(a)] The strong norm $\|.\|_s$ is $\Gamma$-invariant, i.e. $\|\gamma u\|_s=\|u\|_s$
for all $u\in\mathcal{B}$ and $\gamma\in \Gamma$.
\item[(b)] For each fixed $\gamma\in\Gamma$, the linear map $u\mapsto\gamma u$ is
bounded in the weak norm.   (So $\|\gamma u\|_w\le K\|u\|_w$ for all $u\in\mathcal{B}$,
where $K=\|\gamma\|_w<\infty$ is the operator norm of $\gamma$.\footnote{$\|\gamma\|_w=\sup_{u\in\mathcal{B}} \|\gamma u\|_w/\|u\|_w$.})
\item[(c)]
The map $\gamma\to \|\gamma\|_w$ is continuous.
\end{itemize}

\item[(H2)]
There is a family of functions $\Lambda=\{\lambda\}$ of functions
$\lambda:D\rightarrow (0,\infty)$ such that
\begin{itemize}
\item[(a)] $\|u\|_w \leq \| \lambda u \|_s \leq \|u\|_s$, and 
\item[(b)] For all $M>0$ and $\epsilon>0$, there exists $\lambda\in\Lambda$ 
such that
\[ 
\|u\|_s\le M\enspace\text{implies that}\enspace\| (1-\lambda) u\|_w < \epsilon.
\]
\end{itemize}
\item[(H3)]
For all $u\in \mathcal{B}$, $v\in\clos_w(\Gamma u)$, $\lambda\in\Lambda$ 
and $\epsilon>0$, there exists $\gamma\in\Gamma$ such that
\[
\| \lambda (\gamma u-v)\|_s<\epsilon.  % \label{eqlambda3}
\]
\end{itemize}

We give an example of a setting of weighted norms where these hypotheses
can be verified in Section~\ref{sec_rnexample} below.
In our example, $\mathcal{B}$ is a Banach space under the strong norm but not
the weak norm.  This is the typical situation for the applications we have in
mind, but completeness is not used in this paper.

\subsection{Dynamics on $\mathcal{B}$}

Suppose that we have an evolution given by a semiflow $\Phi_t:\mathcal{B}\rightarrow \mathcal{B}$ 
such that
$$
\Phi_t\circ \Phi_s=\Phi_{s+t},~~\Phi_0={\rm Id}.
$$
We suppose that $\Phi_t$ is continuous in both the weak norm $\|.\|_w$ 
and the strong norm $\|.\|_s$ on the function space $\mathcal{B}$.
We suppose also that $\Phi_t$ is $\Gamma$-equivariant.

We say $u\in \mathcal{B}$ has symmetry or {\em isotropy }
$\Sigma_u= \{ \gamma\in \Gamma~:~\gamma u= u \}$ (see for example \cite{GSS88}).
A subgroup $\Sigma$ of $\Gamma$ is {\em cocompact} if the coset space 
$\Gamma/\Sigma$ is compact.

We say $u_0\in\mathcal{B}$ is a {\em relative equilibrium} if
$\Phi_t(u_0)=u(t)=e^{\eta t}u_0$ for some $\eta\in L\Gamma$ which is called the {\em drift}
of $u_0$. Note that for any $\gamma\in\Gamma$, $\gamma u_0$ is also
a relative equilibrium with drift $\eta_\gamma$, where $\eta_\gamma$ is given
by the adjoint action of $\gamma$ on $\eta$ (i.e. 
$e^{\eta_\gamma t}=\gamma e^{\eta t}\gamma^{-1}$ for all $t>0$.)

\begin{rmk}
Associated to the relative equilibrium $u(t)=e^{\eta t}u_0$
is the closed Lie subgroup
$$
K(\eta)=\overline{\{\exp \eta t~:~t\in\R\}}.
$$
Recall that for $\Gamma$ compact and
any $\eta$ the group $K(\eta)$ is a torus and for generic
$\eta$ this torus is maximal~\cite{Field80,Kru00}. 
For $\Gamma$ non-compact, $K(\eta)$ is either isomorphic to
$\R$ or to a torus; generically $K(\eta)$ is either
a maximal torus or $\R$, see \cite{AshMel97}.  For $\E(n)$, generically 
$K(\eta)$ is a torus for $n$ even, and generically $K(\eta)\cong\R$ for $n$ odd.
\end{rmk}

\subsection{An example in $\R^n$ satisfying (H1)--(H3)}\label{sec_rnexample}

Let $\mathcal{B}$ be a set of functions $u:\R^n\to\R^s$.
We assume that $\Gamma$ is a closed subgroup of $\E(n)$
acting as Euclidean isometries in the domain variables $\R^n$,
possibly coupled with a norm-preserving action in the range $\R^s$.
More precisely, let $\gamma=(A,a)\in\E(n)=\O(n)\ltimes\R^n$
act on $x\in\R^n$ by $\gamma x=Ax+a$,
and let $\chi:\Gamma\to O(s)$ be an orthogonal action of $\Gamma$ on
$\R^s$.
Then we assume that the action of $\gamma$ on functions $u\subset \mathcal{B}$
is given by 
\[
(\gamma u)(x)=(\chi_\gamma u)(\gamma^{-1}x).
\]

Define the strong norm $\|u\|_s=\sup_x |u(x)|$ and the weak norm
$\|u\|_w = \|\rho u\|_s$ where $\rho(x)=(1-|x|^2)^{-1}$ and 
$|x|^2=x_1^2+\dots+x_n^2$.  We also define the family of weights 
$\lambda_\alpha(x)=\rho(\alpha x)$, $0<\alpha\leq 1$.   

\begin{lemma}
In this setting, hypotheses (H1)--(H3) are satisfied
with $\Lambda=\{\lambda_\alpha:0<\alpha\leq 1\}$.
\end{lemma}

\proof
Let $\gamma\in\Gamma$, $u\in\mathcal{B}$.
It follows from orthogonality of the action $\chi$ on $\R^s$ that
 $|(\gamma u)(x)|=|u(\gamma^{-1}x)|$.
In particular, $\|\gamma u\|_s=\|u\|_s$ proving (H1)(a).

For any $a\in\R^n$, $\lim_{|x|\to \infty}\rho(x+a)/\rho(x)=1$
so we can define $C(a)=\sup_x \rho(x+a)/\rho(x)<\infty$.
Writing $\gamma x=Ax+a=A(x+A^{-1}a)$, we claim that $\|\gamma\|_w=C(A^{-1}a)$.
It suffices to show that $\|\gamma\|_w=1$ for $\gamma x=Ax$
and $\|\gamma\|_w=C(a)$ for $\gamma x=x+a$.

Note that $\|\gamma u\|_w=\sup_x\rho(x)|u(\gamma^{-1}x)|
=\sup_x \rho(\gamma x)|u(x)|$.
If $\gamma x =Ax$, then $\rho(\gamma x)=\rho(x)$ and $\|\gamma\|_w=1$.   
If $\gamma x=x+a$ is a translation, then it follows from the definition of
$C(a)$ that $\|\gamma u\|_w=C(a)$.
This completes the proof of (H1)(b).

It is clear that $C(a)$ depends continuously on $a$ so that
$\|\gamma\|_w$ depends continuously on $\gamma$, proving (H1)(c).

Let $\alpha>0$ and note that for any $x$
$$
| \rho(x) u(x) | \leq | \lambda_\alpha u(x) | \leq |u(x)|
$$
proving (H2)(a).  To verify (H2)(b), suppose that $\|u\|_s<M$ so that
\[% \label{eqestimate1}
\|(1-\lambda_\alpha) u\|_w = \sup_x | \rho(x) (1-\lambda_\alpha(x))u(x)|\le
P_\alpha M
\]
where 
$$
P_\alpha = \sup_r \frac{1}{1+r^2}\,\frac{\alpha^2 r^2}{1+\alpha^2r^2}\le\alpha^2.
$$
Hence $P_\alpha\rightarrow 0$
as $\alpha\rightarrow 0$, and so we can choose $\lambda=\lambda_\alpha$
with $\alpha$ sufficiently small.

For (H3), let $M=\|u\|_s+\|v\|_s$.   Since $v\in\clos_w(\Gamma u)$, 
there is a sequence $\gamma_n$ such that $\gamma_n u -v\to_w 0$.
In particular, $\gamma_n u-v\to0$ pointwise.
Moreover, $\sup_x|\gamma_n u(x)-v(x)|\le M$ and so it is easy to verify
that $\lambda(\gamma_n u -v)\to0$ uniformly for each fixed $\lambda\in\Lambda$
as required.
\qed

\vspace{5mm}

It is routine to extend this result to the case of a $C^k$ norm, $k\ge1$.
As a trivial example of a semiflow that evolves continuously on $\mathcal{B}$, take any
semiflow that evolves continuously according to its local value, i.e. such that
$$
(\Phi_t (u))(x) = F_t(u(x))
$$
where $F_t$ is a continuous semiflow on $\R^s$. 
Less trivial examples are given by solutions of reaction-diffusion systems.

\section{Co-solutions and relative equilibria}\label{seccosolutions}

Suppose that $\Phi_t$ is a semiflow on $\mathcal{B}$ that is continuous in
strong and weak norms satisfying the hypotheses in Section~\ref{secsetting}. 

For continuous action of compact groups, relative equilibria are
compact and hence closed. As noted in \cite{AshMel97}, this is not true for
noncompact groups unless one makes further assumptions.   Generally 
speaking, in the 
situations of interest in this paper, the relative equilibria are
closed in the strong topology but not in the weak topology.

\begin{defn}   
Let $u_0,v_0\in \mathcal{B}$.  We say that $v_0$ is a {\em co-solution} of $u_0$
if $v_0\in \clos_w(\Gamma u_0)$.
\end{defn}

If $u_0,v_0\in\mathcal{B}$ and $u(t)=\Phi_t (u_0)$, $v(t)=\Phi_t (v_0)$ are the
corresponding solutions, then we say that $v(t)$ is a co-solution of
$u(t)$ if $v_0$ is a co-solution of $u_0$.
It follows from weak-continuity of the flow that the property
$v(t)\in\clos_w(\Gamma u(t))$ holds for one value of $t$ if and only if it 
holds for all $t$. Hence the set of co-solutions of any given solution is
also an invariant set. 

\begin{rmk}
(a) Note that $v_0$ being a co-solution for $u_0$ means that one can 
find arbitrarily large patches
of $u_0$ that resemble $v_0$ arbitrarily closely, up to transformation
by elements of $\Gamma$.
\\[.75ex]
(b) Our definition of co-solution is in terms of the weak topology.
We can also define a strong notion of co-solution.
However in many situations of interest, the notion is vacuous.
Indeed, suppose that $u_0$ is a relative equilibrium with isotropy $\Sigma$.
Following~\cite[Definition~5.2]{AshMel97}, we say that a sequence
$\{\gamma_n\}\subset\Gamma/\Sigma$
is an {\em approximate symmetry} of $u_0$ if
$\gamma_n$ has no convergent subsequences and $\|\gamma_nu_0-u_0\|_s\to0$.
If no such approximate symmetries exist, then in the strong topology
$\Gamma u_0$ is a closed submanifold diffeomorphic to $\Gamma/\Sigma$
(see~\cite[Proposition~5.3]{AshMel97}).
\\[.75ex]
(c) %% An isotropy subgroup $\Sigma$ is {\em cocompact} if $\Gamma/\Sigma$
%% is compact.
Arguing as in (b), we note that relative equilibria with
cocompact isotropy subgroup are compact and hence closed in both the
strong and weak topologies.   In particular, such 
relative equilibria cannot have nontrivial co-solutions.
\end{rmk}

It is clear the co-solutions of equilibria are themselves equilibria.
In certain cases, co-solutions of relative equilibria
are also relative equilibria.

\begin{prop} \label{prop-cosols}
Suppose $u_0$ is a relative equilibrium with drift $\xi$. If there is a 
sequence $\gamma_n\in\Gamma$, an $\eta\in L\Gamma$ and a $v_0\in\mathcal{B}$ such that
$\|\gamma_n u_0-v_0\|_w\rightarrow 0$  and
$$
\|(e^{\xi_n t}-e^{\eta t})\gamma_n u_0\|_w\rightarrow 0
$$
as $n\rightarrow \infty$ 
(where $\xi_n=\xi_{\gamma_n}$ is the drift of $\gamma_n u_0$) for all $t>0$, 
then $v_0$ is a relative equilibrium with drift $\eta$.
\end{prop}

\proof
We set $u_n=\gamma_n u_0$ and calculate
\begin{eqnarray*}
\|e^{\eta t}v_0-\Phi_t (v_0)\|_w & = & \|e^{\eta t}v_0-e^{\eta t}u_n+e^{\eta t}u_n -e^{\xi_n t}u_n+e^{\xi_n t}u_n -\Phi_t (v_0)\|_w \\
& \leq & \|e^{\eta t}v_0-e^{\eta t}u_n\|_w+\|e^{\eta t}u_n -e^{\xi_n t}u_n\|_w+\|\Phi_t (u_n) -\Phi_t (v_0)\|_w.
\end{eqnarray*}
In the limit $n\rightarrow\infty$ for fixed $t$, by continuity of the group action the 
first term goes to zero, by the hypothesis the second term goes to zero and by
continuity of the flow the third term goes to zero. Hence 
$$
\Phi_t (v_0)= e^{\eta t} v_0
$$
meaning that $v_0$ is a relative equilibrium with drift $\eta$.
\qed

\vspace{5mm}

\begin{rmk}
(a) A special case where Proposition~\ref{prop-cosols} applies is when $v_0\in\clos_w(\Gamma u_0)$ 
and $v$ has a drift that is in the centre of $\Gamma$. In such a case $\xi_n=\xi$ and we 
can choose $\eta=\xi$ to satisfy the hypotheses.
\\[.75ex]
(b) Another special case satisfying these hypotheses is where $v$ has full symmetry, 
in which case $\eta=0$.
\end{rmk}

\subsection{Examples of co-solutions}

\begin{figure}
\centerline{
\mbox{\epsfig{file=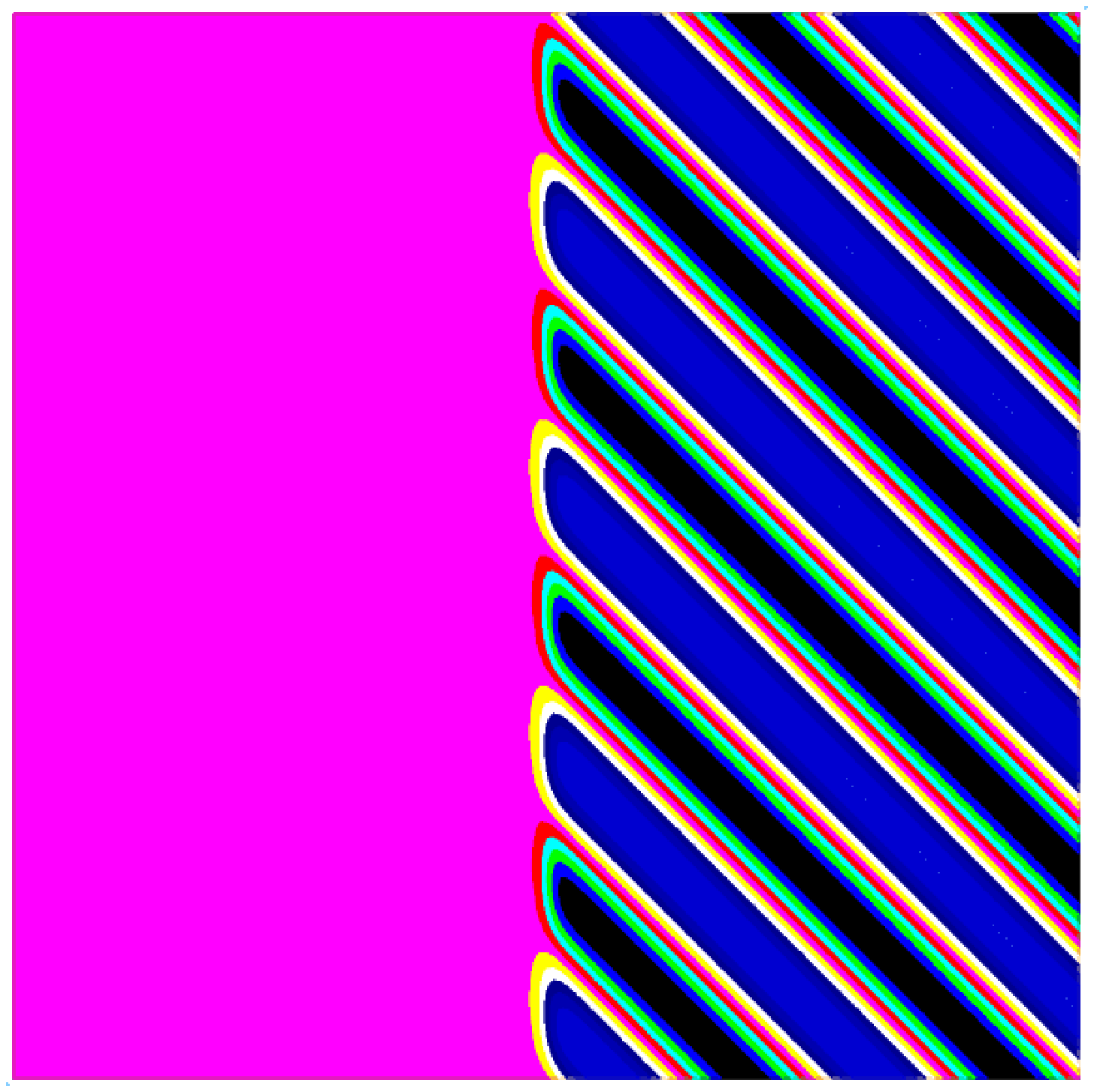,width=8cm,angle=-90}}~
\mbox{\epsfig{file=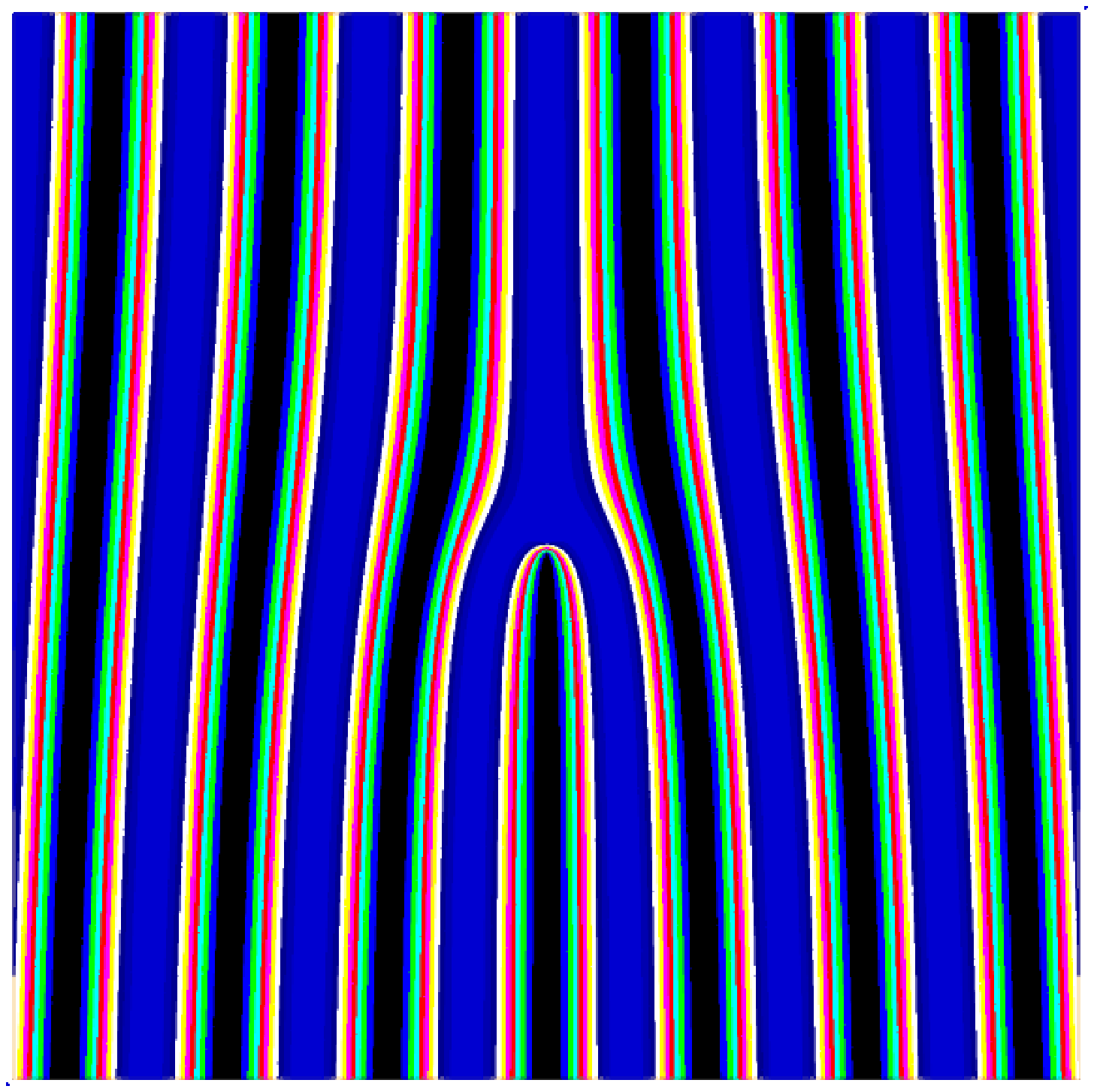,width=8cm,angle=-90}}
}
\center{(a)\hspace{8cm}(b)}
\caption{\label{fig_front}
(a) A relative equilibrium for a Euclidean-equivariant system 
with two co-solutions that are relative equilibria; a uniform state (by taking 
translations to the right) and a family of spatially periodic 
states (by taking translations to the left). (b) 
A relative equilibrium that is a defect of this form has
stripe co-solutions that are relative equilibira on taking translations in any 
direction. In both cases the co-solutions have additional symmetries.
}
\end{figure}

To motivate the results we give a few examples of patterns that have a nontrivial set 
of co-solutions, building on ideas in \cite{Ash00}.  Figure~\ref{fig_front} shows two 
examples. Figure~\ref{fig_front}(a)
shows a front between a spatially periodic
pattern for $x>0$ and a uniform state for $x<0$. If this pattern is a relative
equilibrium for a flow that fits our setting then there are
two families of co-solutions; the uniform pattern for 
$x\to -\infty$ and the periodic pattern for $x\to +\infty$. 
Figure~\ref{fig_front}(b)
shows a defect solution that implies the existence of
stripe solutions as well.  Note that in both cases all co-solutions have
cocompact symmetry, so the co-solutions have no further co-solutions.

\begin{figure}
\centerline{\mbox{\epsfig{file=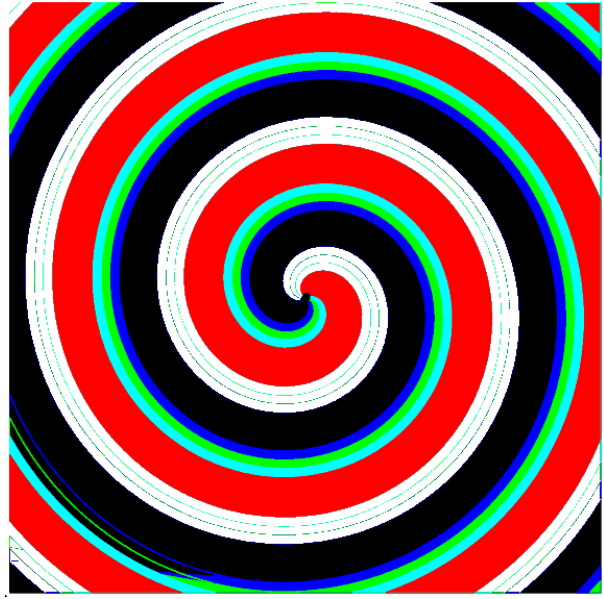,width=8cm,angle=-90}}}
\caption{\label{fig_spiral}
A spiral relative equilibrium for a Euclidean-equivariant system with 
co-solutions given by rolls that drift if the spiral rotates; these are
be found by considering translates of the spiral in any direction.
}
\end{figure}

Another example is illustrated in Figure~\ref{fig_spiral}; this shows one component of
reaction diffusion system with a spiral relative equilibrium that rotate 
anticlockwise. By taking limits of large translations in any direction we obtain
weak co-solutions that are propagating spatially periodic stripe patterns.

\subsection{Symmetries of co-solutions}

\begin{figure}
\centerline{\mbox{\epsfig{file=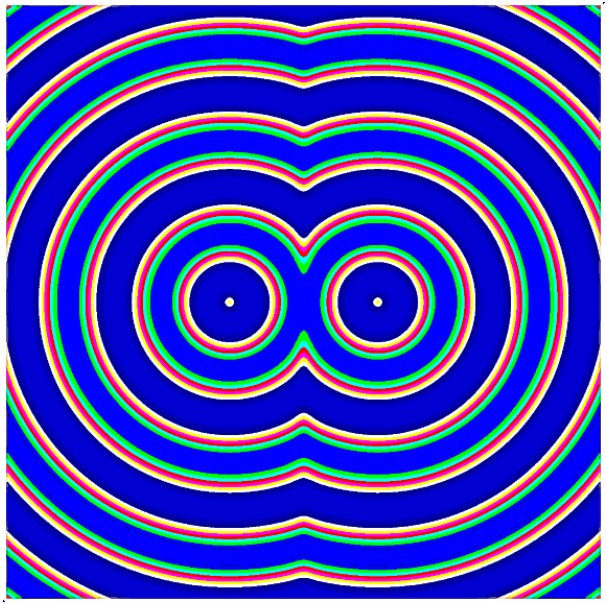,width=8cm,angle=-90}}}
\caption{\label{fig_2target}
Consider an equilibrium for a Euclidean-equivariant system that has two target patterns
anchored near each other. Observe that this has co-solutions that are spatially periodic
stripe patterns; moreover the vertical axis reflection symmetry of the 
original pattern is missing in taking the stripe pattern cosolutions.
}
\end{figure}

In spite of the fact that co-solutions are generated by symmetries of the 
system there is not a simple relationship between the symmetries of a relative
equilibrium $v$ and the symmetries of a co-solution. 
Figures~\ref{fig_front} and \ref{fig_spiral} show cases where the co-solutions
have more symmetry than the original pattern. By contrast Figure~\ref{fig_2target}
has a reflection symmetry in the vertical axis that is missing on the cosolutions obtained
by translating to the left or right. Hence symmetry may be gained or lost in passing 
from a relative equilibrium to a co-solution.

\section{Inheritance of stability}\label{secinheritproof}

Let $u_0\in\mathcal{B}$.  We say that $u_0$ is (Lyapunov) {\em sw-stable} if for 
all $\epsilon>0$ there exists $\delta>0$ such that 
\begin{equation}\label{eqswstab}
\| u - u_0 \|_s < \delta \enspace\text{implies that}\enspace
\|\Phi_t(u)-\Phi_t(u_0)\|_w<\epsilon\enspace\text{for all $t>0$}.
\end{equation}
Similarly, we say $u_0$ is {\em ss-stable} if (\ref{eqswstab}) holds
with $\|.\|_w$ replaced by $\|.\|_s$. This corresponds to the usual 
notion of stability.  Observe that ss-stability clearly implies sw-stability. 

In a similar way one could define ww-stability but this is probably 
too weak to be of use and so we do not discuss it further here.

\begin{prop} \label{prop-sw}
If $u_0\in\mathcal{B}$ is sw-stable (resp.\ ss-stable) then so is 
$\gamma u_0$ for all $\gamma\in\Gamma$.
\end{prop}

\proof
Suppose that $u_0$ is sw-stable and $\gamma\in\Gamma$.
By (H1)(b), $K=\|\gamma\|_w<\infty$.
For any $\epsilon>0$, there exists $\delta>0$ such that
$\|u-u_0\|_s<\delta$ implies that $\|\Phi_t (u)-\Phi_t (u_0)\|_w<\epsilon/K$
for all $t>0$.

We show that $v_0=\gamma u_0$ is sw-stable.
Suppose that $\|v-v_0\|_s<\delta$.   
By (H1)(a), $\|\gamma^{-1}v-u_0\|_s<\delta$.
Hence $\|\Phi_t(\gamma^{-1}v)-\Phi_t (u_0)\|_w<\epsilon/K$.
By (H1)(b) and equivariance of the flow,
\[
\|\Phi_t (v)-\Phi_t (v_0)\|_w\le K\|\gamma^{-1}(\Phi_t (v)-\Phi_t (v_0))\|_w
=K\|\Phi_t(\gamma^{-1}v)-\Phi_t (u_0)\|<K\epsilon/K=\epsilon,
\]
proving that $v_0$ is sw-stable.

The proof that ss-stability of $u_0$ is inherited by $v_0$ is simpler
(with $K=1$) by (H1)(a).
\qed

\vspace{5mm}

\begin{thm} \label{thm-sw}
Suppose that $u_0,v_0\in\mathcal{B}$ and 
$v_0\in\clos_w(\Gamma u_0)$. If $u_0$ is ss-stable then $v_0$ is sw-stable.
\end{thm}

\proof
We prove the statement by contradiction, assuming that $v_0$ is sw-unstable 
and arguing that $u_0$ must be ss-unstable.

Since $v_0$ is sw-unstable, there is an $\epsilon>0$ such that for all 
$\delta>0$ 
we can find a $T>0$ and $v_1$ (both depending on $\delta$) such that
\begin{equation} \label{eq-sw1}
\| v_1-v_0\|_s < \delta,
\end{equation}
but
\begin{equation}\label{eq-sw2}
\|\Phi_T(v_1)-\Phi_T(v_0)\|_w \ge \epsilon.
\end{equation}

By weak continuity of $\Phi_t$, there exists $\eta\in(0,\delta)$ such that
\begin{align}
\|\Phi_T(z)-\Phi_T(v_0)\|_w & < \frac{\epsilon}{3}\quad\text{for all $z\in \mathcal{B}$
with $\|z-v_0\|_w<\eta$}, \label{eq-wc1} \\
\|\Phi_T(z)-\Phi_T(v_1)\|_w & < \frac{\epsilon}{3}\quad\text{for all $z\in \mathcal{B}$
with $\|z-v_1\|_w<\eta$} \label{eq-wc2}.
\end{align}

Set $M=\|u_0\|_s+\|v_1\|_s$.
Then $\|\gamma u_0-v_1\|_s\le M$ for all $\gamma\in\Gamma$ by (H1)(a),
and hence by (H2)(b) there exists $\lambda_0\in\Lambda$ such that
\begin{equation} \label{eq-lambda0}
\|(1-\lambda_0)(\gamma u_0-v_1)\|_w<\eta \quad\text{for all $\gamma\in\Gamma$}.
\end{equation}
Since $v_0\in\clos_w(u_0)$, it follows from (H3) that
there exists $\gamma_0\in\Gamma$ such that
\begin{equation} \label{eq-gamma0}
\|\lambda_0(\gamma_0 u_0-v_0)\|_s<\eta<\delta.
\end{equation}
By hypothesis (H2)(a) and estimate~\eqref{eq-gamma0},
\[
\|\gamma_0 u_0-v_0\|_w\le \|\lambda_0(\gamma_0 u_0-v_0)\|_s<\eta,
\]
so it follows from~\eqref{eq-wc1} that
\begin{equation} \label{eq-est0}
\|\Phi_T(\gamma_0 u_0)-\Phi_T(v_0)\|_w < \frac{\epsilon}{3}.
\end{equation}

Now define
\[
u_1=(1-\lambda_0)\gamma_0 u_0+\lambda_0 v_1.
\]
Then $u_1-\gamma_0 u_0=\lambda_0(v_1-\gamma_0u_0)$ and we compute that
\begin{align*}
\|\gamma_0^{-1}u_1-u_0\|_s & = \|u_1-\gamma_0 u_0\|_s 
= \|\lambda_0(v_1-\gamma_0u_0)\|_s \
\le \|\lambda_0(v_1-v_0)\|_s+\|\lambda_0(\gamma_0 u_0-v_0)\|_s \\
&\le \|v_1-v_0\|_s+\|\lambda_0(\gamma_0 u_0-v_0)\|_s 
 < \delta+\delta= 2\delta,
\end{align*}
where we have used hypotheses (H1)(a) and (H2)(a),
and estimates~\eqref{eq-sw1} and~\eqref{eq-gamma0}.

Moreover, $u_1-v_1=(1-\lambda_0)(\gamma_0 u_0-v_1)$ so 
$\|u_1-v_1\|_w<\eta$
by~\eqref{eq-lambda0}.   It follows from~\eqref{eq-wc2} that
\begin{align} \label{eq-est1}
\|\Phi_T(u_1)-\Phi_T(v_1)\|_w < \frac{\epsilon}{3}.
\end{align}
Writing 
\[
\Phi_T(u_1)-\Phi_T(\gamma_0 u_0) 
= [\Phi_T(u_1) - \Phi_T(v_1)] 
+ [\Phi_T(v_1) - \Phi_T(v_0)] 
+ [\Phi_T(v_0) - \Phi_T(\gamma_0 u_0)]
\]
we have
\begin{align*}
& \|\Phi_T(\gamma_0^{-1}u_1)-\Phi_T(u_0)\|_s  =
\|\Phi_T(u_1)-\Phi_T(\gamma_0 u_0)\|_s \\ &\qquad\ge
\|\Phi_T(v_1)-\Phi_T(v_0)\|_w -
\|\Phi_T(u_1)-\Phi_T(v_1)\|_w -
\|\Phi_T(\gamma_0 u_0)-\Phi_T(v_0)\|_w  \\
&\qquad\ge \epsilon - \frac{\epsilon}{3}-\frac{\epsilon}{3} =\frac{\epsilon}{3},
\end{align*}
where we have used hypothesis (H1)(a),
$\Gamma$-equivariance of $\Phi_T$, and
estimates~\eqref{eq-wc2}, \eqref{eq-est0} and \eqref{eq-est1}.

Summarizing, we have shown that
there is an $\epsilon>0$ such that for all $\delta>0$ there is a $T>0$
and a $w = \gamma_0^{-1}u_1$ such that
$$
\| w-u_0\|_s<2\delta \quad\text{and}\quad
\|\Phi_T(w)-\Phi_T(u_0) \|_s \ge \frac{\epsilon}{3}
$$
giving ss-instability of $u_0$ and the proof is complete.
\qed

\vspace{5mm}

In certain situations we obtain a more powerful result, namely in the presence
of an additional hypothesis:

\begin{itemize}
\item[(H4)]
For any $\epsilon>0$, there exists $\delta>0$ such that for any $u\in\mathcal{B}$
\[
{\textstyle\sup_{\gamma\in\Gamma}}\|\gamma u\|_w<\delta\enspace\text{implies that}
\enspace \|u\|_s<\epsilon.
\]
\end{itemize}

Note that hypothesis (H4) clearly holds for the setup in 
Section~\ref{sec_rnexample} 
where we have $\|u\|_s=\sup_{\gamma\in\Gamma}\|\gamma u\|_w$.

\begin{lemma} \label{lem-ss}
Suppose that (H1-H4) hold and that $u_0\in\mathcal{B}$ has cocompact 
isotropy $\Sigma$.
Then $u_0$ is ss-stable if and only if $u_0$ is sw-stable.
\end{lemma}

\proof
We prove the nontrivial direction, namely that sw-stability implies 
ss-stability.
Let $\epsilon>0$
and choose $\delta>0$ as in (H4).
By Proposition~\ref{prop-sw}, $\gamma u_0$ is sw-stable for all
$\gamma\in\Gamma$.  Hence, for each $\gamma$, there exists 
$\eta=\eta(\gamma)>0$ such that
$\|u-\gamma u_0\|_s<\eta$ implies that
$\|\Phi_t(u)-\Phi_t(\gamma u_0)\|_w<\delta$ for all $t>0$.
By the proof of Proposition~\ref{prop-sw}, we can take
$\eta(\gamma)=\eta(1_\Gamma)/\|\gamma\|_w$ where $1_\Gamma$ is the identity
element in $\Gamma$.
By (H1)(c), $\eta(\gamma)$ depends continuously
on $\gamma$.
Clearly $\eta(\gamma)$ can be chosen to be constant on $\Sigma$-cosets.
 Since $\Gamma/\Sigma$ is compact,
it follows that $\eta>0$ can be chosen independent of $\gamma$.

Suppose that $\|v-u_0\|_s<\eta$ and let $\gamma\in\Gamma$.   
By hypothesis (H1)(a),
$\|\gamma v-\gamma u_0\|_s<\eta$, so by the above argument with
$u=\gamma v$ we have 
$\|\Phi_t(\gamma v)-\Phi_t(\gamma u_0)\|_w<\delta$ for all $t>0$.  
By equivariance, 
$\|\gamma(\Phi_t(v)-\Phi_t(u_0))\|_w<\delta$ for all $t>0$ and all $\gamma\in\Gamma$.
By (H4), we deduce that $\|\Phi_t(v)-\Phi_t(u_0)\|_s<\epsilon$ for all $t>0$
and so $u_0$ is ss-stable.
\qed

\vspace{5mm}

Combining Theorem~\ref{thm-sw} and Lemma~\ref{lem-ss} we have:

\begin{thm} \label{thm-ss}
Suppose that $u_0,v_0\in\mathcal{B}$ and
$v_0\in\clos_w(\Gamma u_0)$.   Suppose further that $v_0$ has
cocompact isotropy. If $u_0$ is ss-stable then $v_0$ is also ss-stable.
\end{thm}

The means, for example, that the existence of an ss-stable relative equilibrium of
the form in Figure~\ref{fig_front}(a) implies that both the uniform and the stripe
co-solutions are ss-stable. Similarly if the spiral solution in Figure~\ref{fig_spiral}
is ss-stable then the `far field' roll solutions are are ss-stable.

\section{Discussion} \label{secdiscuss}

We give a novel way of trying to
understand the qualitative behaviour of dynamics on unbounded domains. There
is clearly a great deal more that can be investigated by making use of a strong
and a weak norm that satisfy assumptions such as (H1-H4) to a co-solution.

One direction that seems worth pursuing is the generalisation
to transients. In particular initial conditions may converge
to relative equilibria in a weak sense, and this gives further predictions 
for the existence of co-solutions (see for example the spiral wind-up discussed 
in~\cite{Ash00}); we note that the results in Section~\ref{secinheritproof} apply
equally for solutions and co-solutions that are not relative equilibria.

In another direction, the results above are purely `topological' in nature and do not
attempt to understand the smooth dynamics. This setting may give a way to obtain
results that relate smooth dynamical properties such as spectral stability to topological
properties such as ss- and sw-stability and indeed to understand what qualitative
ingredients a bifurcation theory for such systems should have.

Similarly it would be interesting to discuss asymptotic stability as well as Lyapunov stability
in the setting; we observe that similar to the case for Lyapunov stability there will
be several inequivalent notions of asymptotic stability depending on choice
of norm.

Finally, we remark that there are situations where a flow that
is continuous in the weak and strong topology becomes continuous in only
the strong topology due to the appearance of mean flow effects \cite{Sch&al}. 
At this point, certain of our hypotheses are violated, and it would be
interesting to understand how this impacts on the  relationship between local and
global dynamics.

\paragraph{Acknowledgements}
This research was supported in part by EPSRC research grant number GR/S31662 (PA)
and by a Leverhulme Fellowship (IM).

\bibliographystyle{plain}

\end{document}